\documentclass[pss,fleqn,embeddedheads]{w-art}
\usepackage{times}
\usepackage{w-thm}
\usepackage{psfrag}
\usepackage[]{graphicx}
\begin{document}
\DOIsuffix{theDOIsuffix}
\Volume{XX}
\Issue{1}
\Copyrightissue{01}
\Month{01}
\Year{2004}
\pagespan{1}{}





\title{Kinematics of Persistent Currents in Nanotubes and Tori}
\subtitle{Time-reversal symmetric gauge field and its application}


\author[K. Sasaki]{Ken-ichi Sasaki\footnote{Corresponding
     author: e-mail: {\sf sasaken@imr.tohoku.ac.jp}}
     and Yoshiyuki Kawazoe}
\address{Institute for Materials Research, Tohoku University, 
Sendai 980-8577, Japan}
\begin{abstract}
 The basic properties of conducting electrons in carbon
 nanotubes are reviewed from a theoretical perspective, and studies performed on
 persistent currents in toroidal carbon nanotubes and on the local energy gap in
 deformed nanotubes are reported on.
\end{abstract}
\maketitle                   




\renewcommand{\leftmark}
{K. Sasaki and Y. Kawazoe: Kinematics of Persistent Currents in Carbon
Nanotubes and Tori}

\section{Introduction}

The Aharonov-Bohm (AB) effect shows the importance of the electromagnetic
gauge field in quantum mechanics and the fact that a single electron has a
fundamental unit of magnetic flux $\Phi_0 = 2\pi/e$ (units for which 
$\hbar = c =1$ are used here).
One direct consequence of the AB effect in solid state physics
is persistent currents~\cite{BIL,Webb,Imry}. 
Persistent currents are driven by an external gauge field produced when
a magnetic flux penetrates a hollow area of a system. 
The external gauge field breaks the time-reversal symmetry and can
induce equilibrium persistent currents in the ground state.
In a previous study~\cite{SK-pc}, a twist-induced gauge field was 
proposed that also affects persistent currents but still preserves the
time-reversal symmetry.
This gauge field is useful in analyzing persistent currents in toroidal
carbon nanotubes as well as in other systems~\cite{SKS-fr}. 
On the other hand, the effect of lattice deformation in carbon nanotubes 
on conducting electrons was studied, and it was found that electrons near
the Fermi level are affected by a gauge field produced by the
deformation~\cite{SKS-def}  
(called the ``deformation-induced gauge field'').
Because a deformation depends on position, the deformation-induced
gauge field is also position dependent.
The purpose of this paper is to show that the twist-induced gauge field
corresponds to a constant part of the deformation-induced gauge field, 
meaning that a local extension of the twist-induced gauge field
appears naturally in the low energy dynamics of conducting electrons in
a deformed carbon nanotube. 
The gauge field presented here provides a unified way of examining the effect
of local lattice deformation in carbon nanotubes on the local energy gap.
At the conclusion of this paper, some problems associated with the
above-mentioned gauge field are discussed.

\section{Basics of Carbon Nanotubes}

In this section, the basic ideas used to describe the quantum
mechanical behavior of conducting electrons in carbon
nanotubes~\cite{SDD,SKS-def} are explained.  
These ideas underlie the work presented in subsequent sections.

\begin{SCfigure}[4][h]
 \psfrag{x}{$x$}
 \psfrag{y}{$y$}
 \psfrag{T1}{$a_1$}
 \psfrag{T2}{$a_2$}
 \psfrag{u1}{\small $R_1$}
 \psfrag{u2}{\small $R_2$}
 \psfrag{u3}{\small $R_3$}
 \psfrag{uc}{Unit Cell}
 \includegraphics[width=.53\textwidth]{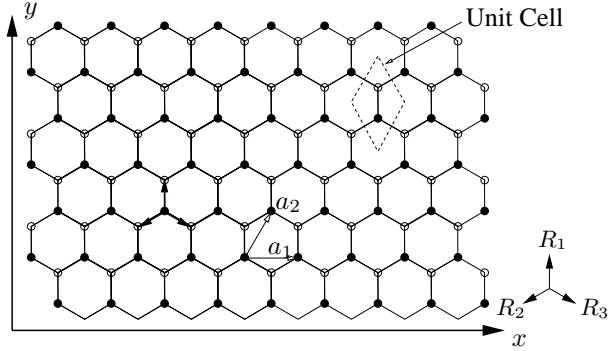}
 \caption{Structure of a honeycomb lattice having the two symmetry
 translation vectors $a_1 = \sqrt{3} a_{\rm cc} e_x$ and 
 $a_2 = (\sqrt{3}/2)a_{\rm cc}e_x + (3/2)a_{\rm cc}e_y$, where 
 $e_x$ and $e_y$ are unit vectors in the direction of $x$ and $y$, and
 $a_{\rm cc}$ is the nearest-neighbor bond length.
 The black (white) circles indicate the A (B) sublattices.
 Vectors $R_a$ $(a = 1,2,3)$ point to the nearest-neighbor sites
 of an A-site.
 They are given by $R_1 = a_{\rm cc} e_y$, 
 $R_2 = -(\sqrt{3}/2) a_{\rm cc} e_x -(1/2) a_{\rm cc} e_y$ and 
 $R_3 = (\sqrt{3}/2) a_{\rm cc}e_x - (1/2) a_{\rm cc} e_y$.} 
 \label{fig:lattice}
\end{SCfigure}

A carbon nanotube is a graphene sheet folded into a cylinder, and
graphene consists of many hexagons (in a honeycomb lattice) whose vertices
are occupied by carbon atoms. Each carbon atom supplies one conducting
electron that determines the electrical property of the carbon nanotube.
The graphene lattice structure is illustrated in
Fig.~\ref{fig:lattice}.
There are two symmetry translational vectors ($a_1$ and $a_2$ satisfying
$|a_1|^2 = |a_2|^2 = 2 a_1 \cdot a_2$) on the planar honeycomb lattice.
The lattice structure of a carbon nanotube can be specified by
two mutually perpendicular vectors, the chiral $C_h$ and the
translational $T$, defined respectively by
$C_h \equiv n a_1 + m a_2$ and $T \equiv p a_1 + q a_2$.
Here, $C_h$ fixes the lattice structure around the axis and is denoted simply
by $C_h = (n,m)$. 
The corresponding wave vector $k$ can be decomposed as $k= \mu_1 k_1 +
\mu_2 k_2$, where $\mu_1$ and $\mu_2$ are integer coefficients of
vectors $k_1$ and $k_2$ satisfying the periodic boundary conditions
$C_h \cdot k_1 = 2\pi$, $C_h \cdot k_2 =0$, $T \cdot k_1 =0$ and $T
\cdot k_2 = 2\pi$.
Thus, $C_h$ and $T$ specify the {\it kinematics} of the electron.

A nontrivial property of a single electron wave function in a honeycomb
lattice is that the wave function has two components. 
This is because two sites exist in the unit cell, depicted as circles
that are black (A-site) or white (B-site) in Fig.~\ref{fig:lattice}. 
The two-component wave function is written as
\begin{align}
 | \Psi \rangle = 
 \begin{pmatrix}
  \Psi_A \cr \Psi_B
 \end{pmatrix}.
\end{align}
It is noted that this two-component nature does not relate to electron
spin.
However, there is a (theoretical) similarity between them such that the
two-component nature is sometimes referred to as a ``pseudo-spin.''
For example, when an electron takes a turn (around the Fermi point) in
k-space through the impurity scattering, the wave function of the
electron takes the (Berry) phase shift of $\pi$ resulting in an absence
of back scattering~\cite{ANS}.
The sign change of the wave function is similar to that of spin
rotation in space.

The Hamiltonian of this system ({\it dynamics}) is thus given as a $2
\times 2$ Hermite matrix.
We consider a general form of the Hamiltonian: 
\begin{align}
 {\cal H} | \Psi \rangle =  
 \begin{pmatrix}
  {\cal H}_{AA} & {\cal H}_{AB} \cr
  {\cal H}_{BA} & {\cal H}_{BB}
 \end{pmatrix}
 \begin{pmatrix}
  \Psi_A \cr \Psi_B
 \end{pmatrix},
\end{align}
where ${\cal H}_{AA}$ is the probability amplitude from an A-site to the
same site, and ${\cal H}_{AB}$ is that from an A-site to the nearest
B-site. 
We assume the nearest-neighbor tight-binding Hamiltonian ${\cal H}_0$,
with hopping integral $V_\pi$, given by
\begin{align}
 {\cal H}_0 = \sum_{a=1,2,3} \sum_{i \in A} V_\pi a_{i+a}^\dagger a_i + h.c.,
\end{align}
where A denotes an A-sublattice, $a_i$ and $a_i^\dagger$ are canonical
annihilation-creation operators of the electron at site $i$ satisfying
the anticommutation relation $\{ a_i,a_j^\dagger \} = \delta_{ij}$, and
site $i+a$ indicates nearest-neighbor sites ($a=1,2,3$) of site
$i$. 
Here, ${\cal H}_{AA}$ and ${\cal H}_{BB}$ are neglected because they are assumed
to be equivalent and to have a constant value, meaning that they do not affect
the energy spectrum but only shift the origin of the energy.
By means of the Bloch theorem, we obtain ${\cal H}_{AB} = {\cal
H}_{BA}^* = V_\pi \sum_a e^{ik\cdot R_a}$.
Here, $R_a$ ($a=1,2,3$) are vectors pointing to the nearest neighbor of
an A-site (see Fig.~\ref{fig:lattice}).
By diagonalizing the Hamiltonian, we obtain the energy eigenvalue as a
function of $k$ as 
$E(k) = \pm V_\pi \left| \sum_{a=1,2,3} e^{ik \cdot R_a} \right|$,
where the negative sign indicates the valence energy band, and the positive
sign denotes the conduction band.
It is crucial to know that there are two positions in k-space where
$E(k)$ vanishes. 
They are located at $\pm K$ where $K = (2b_1 + b_2)/3$ and called the K
and K' points.
Here, $b_1$ and $b_2$ are reciprocal lattice vectors of
the graphene defined by $a_i \cdot b_j = 2\pi \delta_{ij}$ ($i,j=1,2$).

\begin{figure}[htbp]
 \begin{center}
  \psfrag{k2}{$k_2$}
  \psfrag{k3}{$k_2(k'_2)$}
  \psfrag{k1}{$k_1(k'_1)$}
  \psfrag{kd}{$k_2'$}
  \psfrag{k4}{$k_1$}
  \psfrag{k5}{$k'_1$}
  \psfrag{KS}{k-space}
  \psfrag{a}{(a)}
  \psfrag{b}{(b)}
  \includegraphics[scale=0.5]{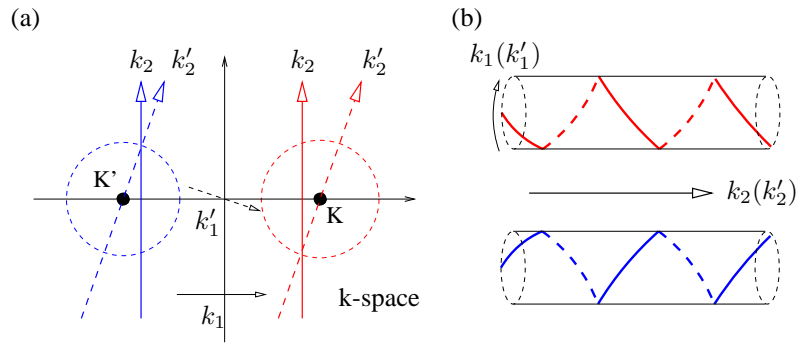}
 \end{center}
 \caption{(a) In k-space, there are two Fermi points (the K and K' points),
 indicated by solid circles, where the valence and conduction bands touch.
 The chiral vector specifies the lines on which energy eigenstates are
 located.
 If lines crossing the Fermi points (dashed arrows) exist, then the
 energy band gap vanishes and the system exhibits metallic
 properties.  
 Otherwise, the system is semiconducting (solid arrows).
 (b) Illustration of low energy modes. 
 The low energy electrons around the K-point (K'-point) generally
 rotate around the axis in the clockwise (counterclockwise)
 direction.
 The lines for armchair $(C_h=(n,n))$ nanotubes correspond to the
 horizontal axis in k-space; therefore, the low energy electrons are
 always classified as non-rotating states.
 }
 \label{fig:motion}
\end{figure}

In the following, metallic or semiconducting behavior dependent on the
lattice structure around axis $C_h$ as exhibited by the carbon nanotube
is shown.
Assuming a nanotube with long length, that is $|T| \gg |C_h|$, then
lines can be drawn in k-space indicating that the energy eigenstates are
located on the lines (see Fig.~\ref{fig:motion}).
The energy gap of the system is determined by the configuration of the
line located nearest to the Fermi points. Here, the electrons
in the lines (the energy bands of the lines) are called ``low-energy
electrons'' (low energy modes).
When the line cross the K-point, the system exhibits metallic properties
and otherwise semiconducting properties. 
This is because the conduction band and valence
band touch only at the K and K' points.
It is important to mention that low energy electrons around the Fermi
points generally rotate around the axis as opposed to a naive
consideration.
It is incorrect to think that low energy electrons should always
have non-rotating modes or that the wave function around the axis is constant.
The motion of low energy electrons around the axis depends on the lattice
structure around the axis.
One can verify the relationship between the chiral vector and the motion
of low energy electrons in Table I of reference~\cite{SK-pc}. 

\section{Time-reversal symmetric gauge field}

In the previous section we explained why nanotubes can show either
metallic or semiconducting behavior, one of the most remarkable
properties of nanotubes.
This is due to (1) the periodic boundary condition {\it around} the axis and
(2) the energy dispersion relation of graphene (there are two Fermi points).
Because the periodic boundary condition around the axis strongly affects
the energy band structure, it would be interesting to consider how the
solid state properties of a nanotube depend on the boundary condition
{\it along} the axis.
If the two ends of a nanotube are connected, the boundary condition
along the axis becomes periodic. 
The resultant geometry, a torus, possesses an important structural
degree of freedom called ``twist''~\cite{MS}.
Twist can mix the motions {\it along} and {\it around} the axis,
and is thus expected to affect the solid state properties.
In Section~\ref{sec:pc}, the kinematics of persistent current in a
twisted carbon torus are reported, and it is shown that current
exhibits the special character of tubule geometry.
It is noted that persistent currents in a twisted torus can be well
described by the notion of a twist-induced gauge field~\cite{SK-pc}. 
Except for the kinematics, the effect of a lattice deformation on the 
low-energy modes is crucial for understanding nanotube physics.
In Section~\ref{sec:def}, the effect of lattice deformation on
the energy gap is examined in terms of a deformation-induced gauge 
field~\cite{SKS-def}. 
Throughout this section, the strong interplay between (1) and (2) has been emphasized,
and it has been remarked that both twist- and deformation-induced gauge fields
maintain the time-reversal symmetry of the system.

\subsection{\bf Twist-induced gauge field}\label{sec:pc}

In this section, persistent currents in toroidal carbon nanotubes
are examined.
Because the effect of twist on persistent currents is of interest,
the difference between persistent currents in an untwisted
torus and that in a twisted torus is important.
Moreover, it is not necessary to consider all of the electrons in the
system. An examination of low-energy electrons is sufficient because the persistent
currents relate to the energy spectrum flow at the Fermi level.
Here, the lattice structure of an untwisted carbon nanotorus is specified initially,
then a twisted nanotorus is considered.

The kinematics for an untwisted torus are considered first.
The wave vector of the low energy modes is given by
$k_\text{low}^\pm = \pm \mu_1^0 k_1 + \mu_2 k_2$,
where $+\mu_1^0$ ($-\mu_1^0$) indicates clockwise (counterclockwise)
rotation around the axis, as shown in Fig.~\ref{fig:motion}(b). 
Because of the periodic boundary condition along the axis, we have
$k_\text{low}^\pm \cdot T = 2\pi \mu_2$. 
Next, the kinematics for a twisted carbon torus are considered.
A specific type of torus, obtained by twisting
an untwisted torus (A-type~\cite{SK-pc}), is used for simplicity.
To express a twist degree of freedom, the
translational vector is replaced as $T \to T_w = T + \frac{t}{d} C_h$,
where $t$ and $d$ are integers.
Here, $t$ denotes the twist and $d$ is the greatest common divisor of the two
integers $n$ and $m$.
The wave vector of the low energy modes in the twisted torus is denoted by
$\bar{k}_\text{low}^\pm$ and should satisfy the following periodic
boundary conditions: 
\begin{align}
 \bar{k}_\text{low}^\pm \cdot T_w = 2\pi \bar{\mu}_2, 
 \ \ \ 
 \bar{k}_\text{low}^\pm \cdot C_h = \pm 2\pi \mu_1^0.
\end{align}
To understand the effect of twist, we consider the inner product between
$\bar{k}_\text{low}^\pm$ and $T$ as
\begin{align}
 \bar{k}_\text{low}^\pm \cdot T 
 = 2\pi \bar{\mu}_2 - \frac{t}{d}
 \bar{k}_\text{low}^\pm \cdot C_h
 = 2\pi \left( \bar{\mu}_2 \mp \frac{t}{d} \mu_1^0 \right).
 \label{eq:t-wave-shift}
\end{align}
The twist shifts the wave vector in k-space and depends on the
electron motion around the axis specified by $\pm \mu_1^0$.
It is important to note that the effect of twist can be provided by a
gauge field for an untwisted torus.
Therefore, the kinematics of a twisted torus are equivalent to that of an
untwisted torus in the presence of a twist-induced gauge field, $A^{\rm
twist}$, satisfying $A^{\rm twist} \cdot C_h = 0$ and $A^{\rm twist}
\cdot T = 2\pi \frac{t}{d} \mu_1^0$.
It is also important to mention that the twist-induced gauge field does
not break the time-reversal symmetry, contrary to the
electromagnetic gauge field $A^{\rm em}$. 
This is because $A^{\rm em}$ shifts the wave vector in one direction and
depends not on electron motion but on 
electron charge $e$.
The electromagnetic gauge field leads to the gauge coupling given by
\begin{align}
 k_\text{low}^\pm \to k_\text{low}^\pm - eA^{\rm em}, \ \ \ \ 
 \left( k_\text{low}^\pm - eA^{\rm em} \right) \cdot T 
 = 2\pi \left( \mu_2 - \frac{\Phi_2^{\rm em}}{\Phi_0} \right),
\end{align}
where $\Phi_2^{\rm em} \equiv A^{\rm em}\cdot T$ is the magnetic flux
penetrating the torus and $\Phi_0$ is the fundamental unit of magnetic
flux. 

\begin{figure}[htbp]
 \begin{center}
  \psfrag{I}{$I_{\rm pc}$}
  \psfrag{F}{$\Phi_2^{\rm em}/\Phi_0$}
  \psfrag{I+}{$+\mu_1^0$ mode}
  \psfrag{I-}{$-\mu_1^0$ mode}
  \psfrag{a}{(a)}
  \psfrag{b}{(b)}
  \psfrag{c}{Add a Twist}
  \psfrag{d}{(c)}
  \psfrag{Total}{Total Current}
  \psfrag{0}{\tiny 0}
  \psfrag{1}{\tiny 1}
  \includegraphics[scale=0.25]{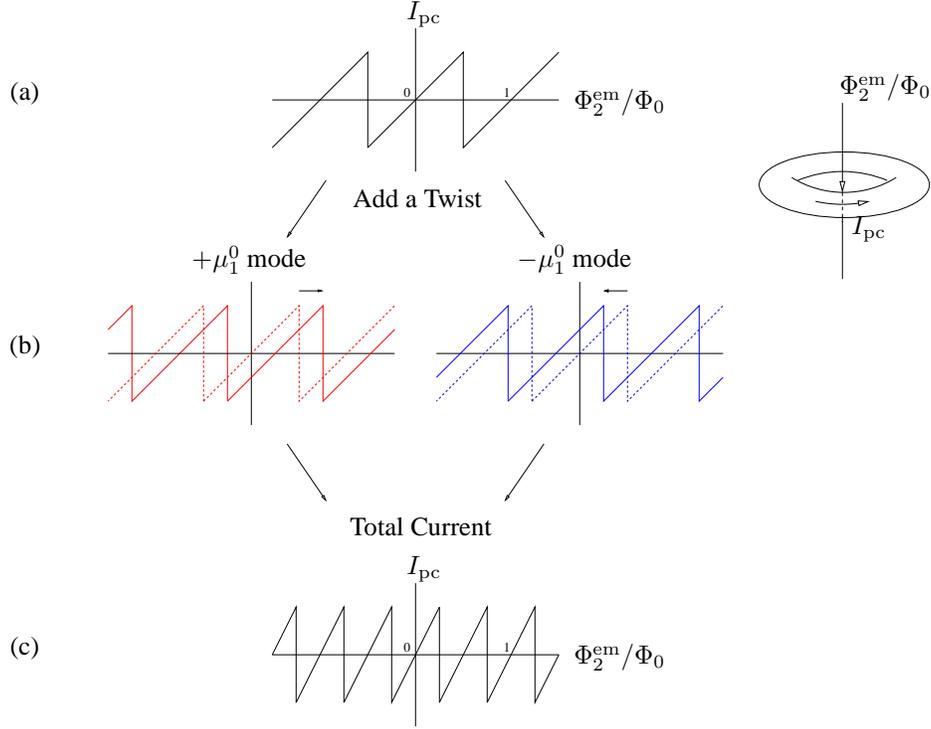}
 \end{center}
 \caption{(a) Sawtooth curve of a persistent current ($I_{\rm pc}$) in
 an untwisted (m-class~\cite{SK-pc}) torus as a function of magnetic
 flux.
 The period of the curve is given by the flux quantum $\Phi_0$.  
 (b) The twist shifts curves in the positive and negative directions 
 depending on the clockwise and counterclockwise electron
 motion around the axis. (c) The total persistent current is given
 by the sum of the two currents.}  
 \label{fig:twist-pc}
\end{figure}

Just as persistent current is induced by magnetic flux $\Phi_2^{\rm
em}$ or $A^{\rm em}_2$, the twist-induced gauge field $A^{\rm twist}$
also affects persistent current.
Sawtooth curves for the K-point and K'-point persistent currents are shifted
in different directions by $A^{\rm twist}$ because of the $\pm$ sign in
Eq.~(\ref{eq:t-wave-shift}).
Suppose the persistent current in an untwisted torus is given by the
usual sawtooth curve as a function of $\Phi_2^{\rm em}$.
In this case, persistent currents of the K-point and K'-point are equal,
and the amplitude of total current is twice that of the K-point only
(Fig.~\ref{fig:twist-pc}(a)). 
The twist shifts the curves in the positive and negative directions
depending on electron motion around the axis as shown in
Fig.~\ref{fig:twist-pc}(b).
The total persistent current is the sum of these two currents, giving
a generally complicated resulting current.
Fig.~\ref{fig:twist-pc}(c) depicts a special case where the period
of the current becomes $\Phi_0/2$.

\subsection{\bf Deformation-induced gauge field}\label{sec:def}

We consider the effect of a lattice deformation on the low energy
modes.
The Hamiltonian of an undeformed tube can be decomposed into two parts
at low energy as 
${\cal H}_0 \Longrightarrow {\cal H}_K \oplus {\cal H}_{K'}$, where
${\cal H}_K \equiv v_F \sigma \cdot p$ (${\cal H}_{K'} \equiv v_F
\sigma' \cdot p$) defines the effective dynamics for the K-point
(K'-point) and the corresponding Schr\"odinger equation is given by the
Weyl equation~\cite{Semenoff}.
Here, the Fermi velocity is denoted by $v_F (=3V_\pi a_{\rm cc}/2)$,
$\sigma_i$ are the Pauli matrices and 
$\sigma \cdot p \equiv \sigma_1 p_1 + \sigma_2 p_2$ 
($\sigma' \cdot p \equiv -\sigma_1 p_1 + \sigma_2 p_2 $). 
It is noted that $p$ is the momentum measured from the K-point or
K'-point.
A lattice deformation can be included by adding the deformed Hamiltonian 
${\cal H}_{\rm deform}$ to ${\cal H}_0$ as
\begin{align}
 {\cal H}_{\rm deform} = 
 \sum_{a=1,2,3} \sum_{i \in A} \delta V_a(r_i) a_{i+a}^\dagger a_i +
 h.c., 
\end{align}
where $\delta V_a(r_i)$ is a modulation of the hopping amplitude from
$r_i$ to $r_{i+a}$, induced by the deformation.
The leading order of the total Hamiltonian, ${\cal H}_0 + {\cal H}_{\rm
deform}$, can be expressed at low energy by 
${\cal H}_K = v_F \sigma \cdot \left( p - A \right)$ and 
${\cal H}_{K'} = v_F \sigma' \cdot \left( p + A \right)$,
where $\sigma \cdot A = \sigma_1 A_1 + \sigma_2 A_2$, $A$ is the
vector field defined on the surface and $v_F A(r)$ is given by a linear
function of $\delta V_a(r)$.
It is noted that the deformation can be included as a gauge field
because of the gauge coupling $p \to p \pm A$.
Here, $A$ is called the deformation-induced gauge field, and it does not break
the time-reversal symmetry because the sign in front of $A$ is different
for the K and K' points.
It is also important to note that this sign characteristic is equivalent to
$A^{\rm twist}$, but in this case $A$ can be position dependent.

\begin{figure}[htbp]
 \begin{center}
  \psfrag{a}{(a) A nanotube is deformed by $C_{60}$}
  \psfrag{b}{(b) Deformation-induced gauge field $A$}
  \psfrag{d}{(d) Energy gap}
  \psfrag{c}{(c) Deformation-induced magnetic field}
  \psfrag{A}{$(A_1,A_2)$}
  \psfrag{x}{$x_2$}
  \psfrag{g}{$E_{\rm gap}(x_2)$}
  \psfrag{p}{$B_3$}
  \psfrag{de}{$C_{60}$}
  \psfrag{T}{$\Phi_1(x_2)$}
  \psfrag{t}{$x_2$}
  \psfrag{T'}{$\Phi_1(x'_2)$}
  \psfrag{t'}{$x'_2$}
  \includegraphics[scale=0.3]{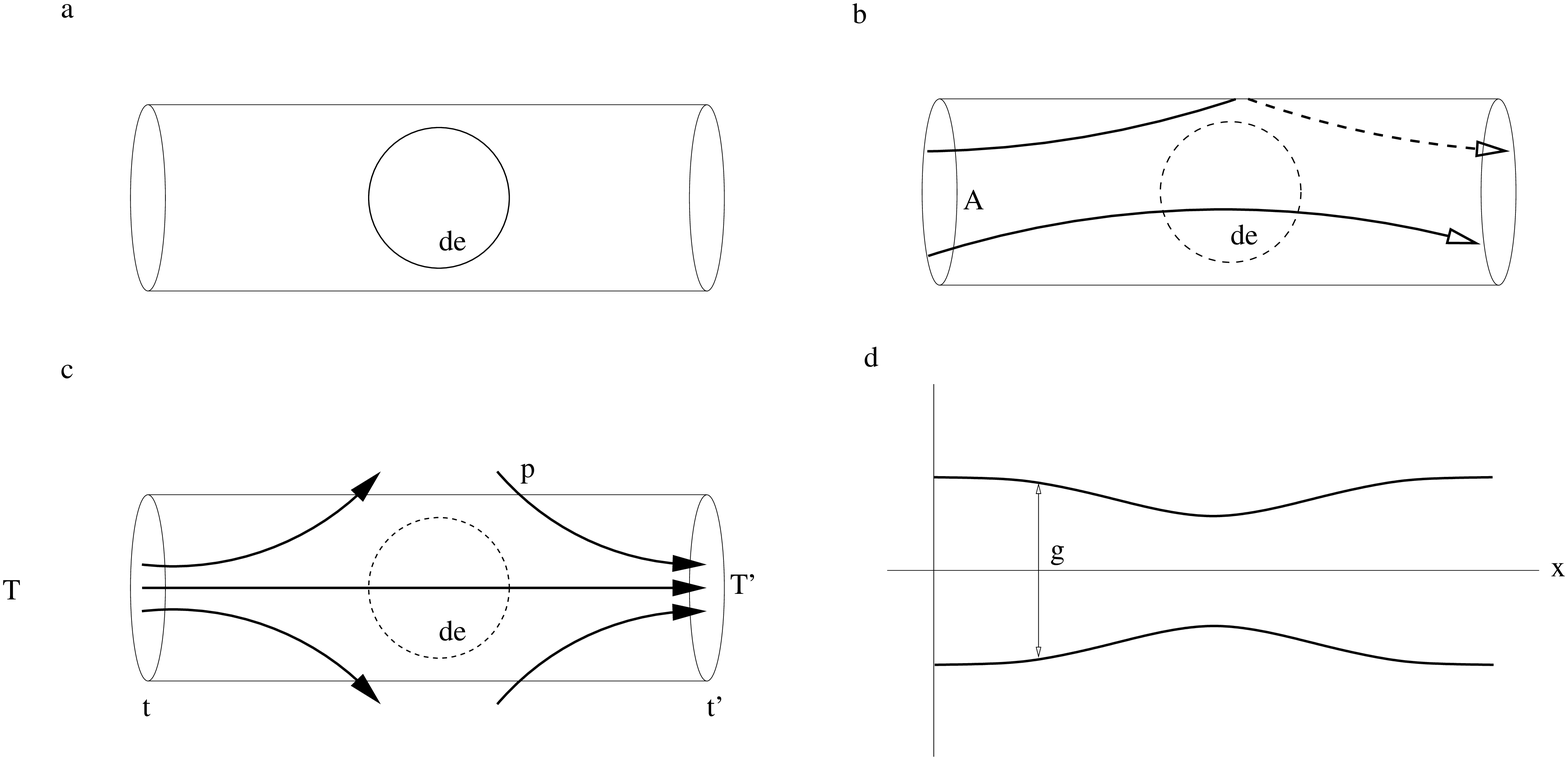}
 \end{center}
 \caption{
 A deformed carbon nanotube (a) can be thought of as an undeformed 
 nanotube in the presence of a deformation-induced gauge field (b).
 The relationship between (c) a deformation-induced magnetic field $B_3$
 (d) the corresponding local energy gap. 
 It is noted that this is not the usual electromagnetic field
 ($\Phi_1^{\rm em} = 0$).} 
 \label{fig:field}
\end{figure}

The effect of $A$ on the energy gap is now considered.
For this purpose, ${\cal H}_K$ is considered in the presence of the
electromagnetic gauge field~\cite{AA,Kono}. 
The Hamiltonian of the system is given by
${\cal H}_K = v_F \sigma \cdot \left( p - A - eA^{\rm em} \right)$,
which indicates that the low energy electrons around the K-point are affected by
the gauge field of $A + e A^{\rm em}$.
The magnetic flux parallel to the tube axis is given by the sum of a
(true) magnetic flux 
$\Phi^{\rm em}_1 \equiv \oint dx_1 A^{\rm em}_1$ 
and a magnetic flux produced by the deformation-induced gauge field
$\Phi^{\rm deform}_1(x_2) \equiv \oint dx_1 A_1(x_1,x_2)$
as $\Phi_1(x_2) \equiv \Phi^{\rm deform}_1(x_2) + \Phi^{\rm em}_1$, 
shifting the wave vector through the AB effect.
This shift corresponds to the fact that the lines in
Fig.~\ref{fig:motion}(a) also shift, resulting in a change of the energy
gap~\cite{KM,Ouyang}. 
The important point here is that the change depends on the axis position
$x_2$, because the flux is a function of $x_2$.
Therefore, the energy gap modulates along the axis.
The energy gap modulation can be estimated by 
\begin{align}
 \delta E_{\rm gap} \approx
 2 v_F \frac{2\pi}{|C_h|} \frac{\Phi_1(x_2)}{\Phi_0}.
\end{align}
It is important to note that a magnetic field appears when the
energy gap is locally modulated.
This magnetic field, perpendicular to the tube surface, can be defined by
$B_3 = \epsilon_{ij} \partial_i A_j = \nabla^2 \Psi_b$.
In fact, we have
$\Phi_1(x_2 + \delta x_2) - \Phi_1(x_2) = \delta x_2 \oint dx_1
\partial_2 A_1 = - \delta x_2 B_3$.
A deformed (by $C_{60}$) nanotube and the corresponding local modulation
of energy gap are illustrated in Fig.~\ref{fig:field}.

It has been assumed thus far that the deformation does not mix the K and
K' points. 
A short-distance lattice deformation, or topological defects such as a
pentagon or heptagon, can mix them~\cite{GGV} such that the
deformation-induced gauge field should be generalized.
A non-Abelian extension of the $A$ field can successfully include these
kinds of deformations~\cite{SKS-def}.

\section{Discussion}

The following sections discuss several problems not yet solved but which seem to be 
interesting subjects.

\subsection{\bf Persistent Currents in the Torus}

An interesting theoretical issue is the need to understand persistent currents
in toroidal carbon nanotubes in more detail.
There are several theoretical models for toroidal shapes composed of
carbon atoms, classified into three groups:   
(t1) a small torus consisting of about hundred carbon atoms~\cite{Ito},
(t2) a torus with a large aspect ratio in the presence of
pentagon-heptagon pairs in the tube surface, and
(t3) a torus with a large aspect ratio in the absence of pentagon-heptagon
pairs.
A simple (t3) torus is considered here.
However, even for (t3), a theoretical study of persistent currents in the presence
of a lattice deformation is not complete.
Another interesting issue is clarification of the effects of topological
defects on persistent currents (t2).  

\subsection{\bf Local Energy Gap at the Edge}

\begin{figure}[htbp]
 \begin{center}
  \psfrag{g}{$A_1$}
  \psfrag{c}{$\delta V_1 \ne 0$}
  \psfrag{d}{bond cut}
  \includegraphics[scale=0.5]{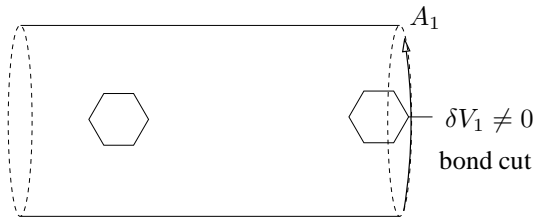}
 \end{center}
 \caption{Illustration of an edge part of zigzag nanotube, $C_h=(n,0)$.
 The hopping integral along the tube axis should be modulated at the
 boundary, giving rise to the deformation-induced gauge field
 around the axis.
 It is noted that a localized edge state exists
 around the zigzag edge whose energy eigenvalue is flat at the Fermi
 energy~\cite{Fujita}.} 
 \label{fig:edge}
\end{figure}

This section discusses the applicability of the local energy gap formula.
Because it is meant for non-mixing cases, it cannot be used in its
present form to estimate the 
local energy gap near topological defects. 
For example, there may be a closed cap structure containing pentagons
at the edge of a tube for which the local energy gap should be considered in
the framework of the non-Abelian deformation-induced gauge
field~\cite{SKS-def}. 
In the absence of topological defects, bonds along the axis direction
cut at the edge, which generally gives a local deformation-induced gauge
field {\it around} the axis $A_1$ (see Fig.~\ref{fig:edge}). 
Hence, it is naturally expected to cause local modulation of the
energy gap at the edge.
It should be noted that the framework for the local energy gap presented
here does not give the wave function of energy eigenstates in a deformed
nanotube. 
This should be clarified in future work.

\begin{acknowledgement}
 We thank R. Saito for discussions.
 K. S. is supported by a fellowship of the 21st Century COE Program
 of the International Center of Research and Education for Materials of
 Tohoku University. 
\end{acknowledgement}

\end{document}